\documentstyle[preprint,aps]{revtex}
\begin{document}
\tightenlines
\draft

\title{$Z \alpha $-expansion for self-energy radiative corrections to parity
nonconservation in atoms}

\author{M.Yu.Kuchiev
\thanks{kuchiev@newt.phys.unsw.edu.au} }
\address{School of Physics,
University of New South Wales,Sydney 2052,  Australia}
\maketitle

\begin{abstract}
The self-energy and vertex QED radiative corrections to the parity
nonconservation (PNC) amplitude in atoms are obtained using 
the perturbation theory in powers of $\alpha Z$.
The calculated linear in $\alpha Z$ term
gives $-0.6 \%$ for the PNC amplitude in Cs. 
The estimated nonlinear terms make corrections larger 
$ -0.9(2)\% $. This result brings the experimental data for 
the $6s-7s$ transition in $^{133}$Cs in agreement with the standard model.

\end{abstract}


This work is inspired by a recently discovered pronounced deviation of
experimental data on parity nonconservation (PNC) in atoms from predictions
of the standard model. The deviation was found in the $6s-7s$ PNC amplitude
in $^{133}$Cs, that was investigated by Bouchiat and Bouchiat
\cite{bouchiat}, Gilbert and Wieman \cite{gilbert_wieman_86}, and, most
recently with the highest precision $0.3 \% $, by Wood {\it et al}
\cite{wood_97}.  The experimental progress was matched by atomic PNC
calculations that are crucial for the analysis of the experimental data.
Accurate calculations of Refs.  \cite{dzuba_89,blundell_92} have been
revisited recently by Kozlov {\it et al} \cite{kozlov_01} and Dzuba {\it et
al} \cite{dzuba_01,dzuba_02}.  Bennett and Wieman \cite{bennett_wieman_99}
compared the theoretical data \cite{dzuba_89,blundell_92} with available
experimental data on dipole amplitudes, polarizabilities and hyperfine
constants for Cs, and suggested that the theoretical error for the PNC
amplitude should be reduced from $1\%$ to $0.4 \%$. This suggestion is
supported by Ref.\cite{dzuba_02} that estimates the theoretical error as
$0.5 \%$.  There have been discovered recently several, previously
neglected phenomena that contribute at this level of accuracy. The Breit
corrections give $-0.6 \%$ as was found by Derevianko \cite{derevianko_00}
and confirmed in \cite{dzuba_harabati_01,kozlov_01}.  Sushkov
\cite{sushkov_01}  proposed that the radiative corrections may be
important. This suggestion is supported by the calculations of Johnson
{\it et al} \cite{johnson_01} who demonstrated that the QED vacuum
polarization gives $0.4 \%$, the value confirmed in
\cite{milstein_sushkov_01,dzuba_01,kf_02}. Ref.  \cite{bennett_wieman_99}
indicates that there is a $2.3 \sigma$ deviation of the weak charge
$Q_{\mathrm W}$ extracted from the atomic PNC amplitude \cite{wood_97} from
predictions of the standard model \cite{groom_00}. More recent analyses
\cite{johnson_01,dzuba_02}, in which the Breit corrections ($-0.6 \%$) and
the  QED vacuum polarization ($0.4 \%$) were included, give similar
deviations  $2.2\sigma$ and $2.0\sigma$ respectively.

We evaluate the self-energy and vertex radiative corrections, called the
e-line corrections below.  They were considered previously by Marciano and
Sirlin \cite{marciano_sirlin_83} and Lynn and Sandars
\cite{lynn_sandars_94} using the plane wave approximation that resulted in
a small value $\sim 0.1 \%$.  An attempt to take into account the strong
Coulomb field in this problem was made in Ref.  \cite{milstein_sushkov_01}
that concluded that the e-line corrections are negligible because they are
not enhanced by large logarithmic factors (which play an important role for
the vacuum polarization, as was demonstrated by Milstein and Sushkov in the
cited paper and confirmed in Ref.  \cite{kf_02}). We show that the strong
Coulomb field makes the e-line corrections  large due to unexpected numerical
interplay.  Our results are in qualitative agreement with conclusions of
recent Ref.\cite{kf_se_02} that demonstrated for the first time that the
e-line corrections are important.  We find quantitative agreement with this
paper as well, the correction $-0.9 \% $ obtained for the $6s-7s$ PNC
amplitude in $^{133}$Cs in the present paper matches the result $-0.7(2) \%
$ of \cite{kf_se_02}. This correction brings the experimental data of Wood
{\em et al } \cite{wood_97} in agreement with the standard model.
The present paper relies on the perturbation theory in powers
of $\alpha Z$.
It should be noted that \cite{kf_se_02} uses very different approach,
based on an identity that expresses the radiative corrections to the PNC
amplitude via corrections to the energy shifts due to the finite nuclear
size that have been calculated previously in
Refs.\cite{johnson_soff_85,blundell_FNS_92,cheng_93,lindgren_93}.

Diagrams (1),(2) and (3) in Fig.1 describe the e-line radiative corrections
to the PNC amplitude in the first order of the perturbation theory in
$\alpha Z$ . The weak PNC interaction between the electron and the nucleus
originates from the Z-boson exchange. It is described by the  Hamiltonian
(relativistic units $\hbar = c = m = 1$ are used, if not stated
otherwise)

\begin{equation}\label{ham} H_{\mathrm PNC} =
(2\sqrt 2)^{-1} \,G_F Q_{\mathrm W}\, \rho(r) \, \gamma_5~,
\end{equation}
where $G_F$ is the Fermi constant, $Q_W$ is the nuclear weak charge, and
the nuclear density for our purposes can be approximated by the delta
function, $ \rho(r) = Z \delta({\bf r}) $.  The perturbation induced by the
Hamiltonian (\ref{ham}) is described by a factor $G_F Q_{\mathrm
W}Z/(2\sqrt 2) \gamma_0 \gamma_5$ that appears for all the diagrams
considered.  Since we are interested merely in a relative contribution of
the corrections, the constant $G_F Q_{\mathrm W}Z/(2\sqrt 2)$ will be
dropped in all formulas below to simplify notation. We use the Feynman
gauge for the photon propagator $D_{\mu\nu}(q) = 4\pi
g_{\mu\nu}/(q^2-\lambda^2)$, and introduce the density
matrix
$\wp = (1/4)\,(\hat p+1)( 1+\hat a\gamma_5)$
that distinguishes the electron state with the given momentum $p_\mu$ and
the polarization four-vector $a_\mu$. The latter can be expressed via the
polarization three-vector ${\mbox{\boldmath$\zeta$}}$, $a_0 =  {\bf p}\cdot
{\mbox{\boldmath $\zeta$} },~ {\bf a} = {\mbox{\boldmath$\zeta$} } +
{\bf p}
({\bf p}\cdot {\mbox{\boldmath$\zeta$} } ) /(1+p_0)$. It suffices to
discuss the forward scattering in which the electron momentum $p_\mu$ and
its polarization $a_\mu$ remain intact, and restrict our consideration to
the limit ${\bf p} \rightarrow 0$ because all important events that
contribute to the process happen at separations $r\simeq 1/m$ that are much
less than the atomic radius. The only pseudoscalar available in this
kinematics is $a_0 = {\bf p} \cdot {\mbox{\boldmath$\zeta$} }$ that should
appear as a common factor for all diagrams for the PNC amplitude. We need
therefore to extract the lowest, linear in ${\bf p}$ term
of the ${\bf p}\rightarrow 0$ expansion.  The analytical expression for
each Feynman diagram can be written in the following form

\begin{equation}\label{ana}
-i \, \alpha  \int
\frac{ d^4q }{ (2\pi)^4} \frac{d^3k}{(2\pi)^3}\,
\frac{4\pi}{q^2-\lambda^2}\,
\frac{\!\!-4\pi Z \alpha }{~{\bf k}^2}  \,
\frac{ {\mathrm Tr}( {\cal N}  \wp )  } {\cal D}~,
\end{equation}
where $q$ is the photon momentum, ${\bf k}$ is the momentum
running over the Coulomb and the PNC weak interactions, ${\mathrm Tr}$ is
the trace over the spinor indexes, and the nominator ${\cal N}$ and
denominator ${\cal D}$ are specific for each diagram. To clarify notation
let us present explicitly the nominators ${\cal N}_1,~{\cal N}_2$ for one
Feynman diagram of the type (1) and one diagram of the type (2) in Fig.1

\begin{eqnarray}\label{nd}
{\cal N}_1 &=& \gamma^\mu (\hat P+1) \gamma_0
(\hat Q+1)\gamma_\mu (\hat R+1) \gamma_0 \gamma_5~,
\\ \nonumber
{\cal N}_2 &=& \gamma^\mu (\hat P+1) \gamma_0 (\hat Q+1)
\gamma_0 \gamma_5 (\hat P+1) \gamma_\mu~,
\end{eqnarray} 
where $
P=p+q,~Q=p+k+q,~R=p+k,~k = (0,{\bf
k})$. The vertexes $\gamma_0$ and $\gamma_0 \gamma_5$ in Eq.(\ref{nd})
originate from the Coulomb and the PNC weak interaction respectively.
Using Eqs.(\ref{iden}) we find from Eq.(\ref{nd}) (and similar expressions
for the Feynman diagrams that are topologically identical to either 
diagram (1), or (2) in Fig.1)

\begin{eqnarray}
\label{avpho}
\sum {\mathrm Tr}({\cal N}_1 \wp) &=&
-8 \,\Big[ \, 2 P_0 \Big( a_0+ (pR)a_0-(aR)p_0 \Big)
-(Q\tilde P \tilde R a) \,\Big]~,
\\ \label{avpho2}
\sum {\mathrm Tr}({\cal N}_2 \wp)
&=& -4 \,[ \, (\tilde Q a) - (P \tilde Q \tilde R a)\, ]~.
\end{eqnarray}
Here $(x y)$ is a scalar product of vectors, $(x y) \equiv x_\mu y^\mu$,
and the symbol $(xyzu)$ stands for the following invariant of four
four-vectors $(xyzu) \equiv (xy)(zu) - (xz)(yu) + (xu)(yz)$. The tilde sign
in Eq.(\ref{avpho}) marks the inversion of a four-vector $\tilde x^\mu
\equiv (x_0,-{\bf x})$.  Summation in  (\ref{avpho}) and (\ref{avpho2})
includes all Feynman diagrams topologically identical to 
diagram (1) and (2) in Fig.1 respectively. The conventional
parameterization permits the denominator for diagram (1) in Fig.1 to be
presented in the form

\begin{eqnarray}\label{ksi}
\frac{ 1 }{(q^2-\lambda^2)\,{\cal D}_1 } = \frac{2}{R^2-1} \int_0 ^1
\frac{1}{D_1^3}
\delta \left( \sum_{i=1} ^3 \xi_i-1 \right) \,\prod_{i=1}^3 d \xi_i~,
\\ \nonumber
D_1 = 
\xi_1(P^2-1) + \xi_2(Q^2-1)+\xi_3(q^2-\lambda^2) ~.
\end{eqnarray}
Using similar presentation for diagram (2) we calculate the integral
over the photon momentum $q$ in Eq.(\ref{ana}) for these two diagrams
with the help of Eqs.(\ref{integrals}).
Expanding the found results up to the first power in ${\bf p}$ we integrate
them over the angular variables $\Omega_{\bf k}$ of the momentum ${\bf k}$.
These straightforward calculations lead to the following representation for
the e-line radiative correction $\delta_{\mathrm e-line}^{\mathrm PNC}$ to
the PNC amplitude

\begin{eqnarray}\label{delta}
& \delta & _{\mathrm e-line}^{\mathrm PNC} = {\cal C}^{\mathrm PNC}\,\alpha^2 Z ~,
\\ \label{F}
&{\cal C}& ^{\mathrm PNC}  = -\frac{8}{3\pi^2} \int_0^\infty
\Big( \,f_1(k^2)+f_2(k^2)+f_3(k^2) \,\Big)\,dk~.
\end{eqnarray}
Note that $\delta_{\mathrm e-line}^{\mathrm PNC}$ is a relative correction,
i.e. the correction divided by the main amplitude (which in our notation is
equal to $(-a_0)$). The integrand in
Eq.(\ref{F}) includes contributions from all diagrams, each $f_i(z),~z=k^2$
originates from the (i)-th diagram in Fig.1. The above described procedure
presents the functions $f_1(z), f_2(z)$ via the two-dimensional
integrals

\begin{eqnarray}\label{f1}
f_1(z) &=& \frac{ 1 }{ z } \int_0^1 du \int_0^u dv \left(
\frac{A}{D}-\frac{B}{D^2}\,z +\frac{C}{u^2}+ 2 \ln \frac{D}{u^2} \right)~,
\\ \label{f2}
f_2(z) &=&  \int_0^1 du \int_0^u dv (u-v)\left(
\frac{E}{2D^2}-\frac{F}{D^3}\,z  - \frac{4}{u^3\,z} -
\frac{3(1-u)}{2D}  \right)~,
\end{eqnarray}
where the symbols $A,B, \dots F$ in the integrands are defined as follows

\begin{eqnarray}\label{ABC}
A &=& (1-u)(4-u+2v)+\Big( \,(1-u)(2-v) + 2 v(1-v) \,\Big)z~,
\\ \nonumber
B &=& (1-u)v \, \Big( \,(1-u)(2+u-2v)+v(1-v)z \,\Big)~,
\\ \nonumber
C &=& 4- (1-u)(4-u+2v)~,
\\ \nonumber
D &=& u^2+v(1-v)z~,
\\ \nonumber
E &=&  (1-u) \,\left[ \, 3\Big( 1 + (1-u)^2 \Big) + v(1-v)z \,\right]~,
\\ \nonumber
F &=&  (1-u) v \, \left[ \,(1-u)^2(1-3v)+1-v +v^2(1-v)z \,\right]~.
\end{eqnarray}
The integration over $u,v$ in (\ref{f1}),(\ref{f2}) arises from  the
integral in Eq.(\ref{ksi}) (and a similar expression for the diagram (2)),
$u=\xi_1+\xi_2,~v=\xi_2$. Diagram (3) in Fig.1 is expressed directly via
the mass operator calculated by Feynman \cite{feynman_49}, resulting in

\begin{eqnarray}\label{f3}
f_3(z) &=& -\frac{1}{2z} (\,2 g(z) + z g'(z)\,) ~,
\\ \label{l}
g(z) &=& \frac{1}{2(1-z)}\left(2-z+\frac{z^2+4z-4}{1-z}\ln z\right)+1~.
\end{eqnarray}
Eqs.(\ref{f1}),(\ref{f2}), and (\ref{f3}) are derived using conventional
renormalization procedure to deal with the logarithmic ultraviolet
divergences. In the Feynman gauge each one of the diagrams in Fig.1
possesses also the infrared singularity $\propto \ln \lambda,~
 \lambda \rightarrow 0 $. It is well known that for problems of the type
considered the infrared divergence should not manifest itself, and,
indeed, we find that it cancels out for the sum of all diagrams.  This
cancellation allows us to derive convergent integral representations
(\ref{f1}),(\ref{f2}), and to express (\ref{f3}) in terms of a well
defined, finite function $g(z)$ (\ref{l}).  Since the functions $A,B,\ldots
F$ in Eq.(\ref{ABC}) are all polynomials, a large number of, probably all,
integrations in (\ref{f1}),(\ref{f2}) can be carried out in an analytical
form. However, our main goal here is a numerical value for the factor
${\cal C}^{\mathrm PNC}$.  We rely, therefore, on numerical calculations
whenever it is more convenient.  Calculating integrals in
Eqs.(\ref{f1}),(\ref{f2}) we find the functions $f_1(z)$ and $f_2(z)$.
\footnote{A simple way provides an expansion of $1/D$ in
Eqs.(\ref{f1}),(\ref{f2}) in powers of $[z v^2/(u^2+vz)]$. For each term in
this series {\em Mathematica} \cite{mathematica} reliably performs
integrations in (\ref{f1}),(\ref{f2}) in an analytical form. The procedure
rapidly converges.} Combining them with $f_3(z)$ from (\ref{f3}) we find
the function $f(z)=\sum_{i=1}^3 f_i (z) $ in the integrand in Eq.(\ref{F}).
Its asymptotes are

\begin{equation}\label{asym}
f(z) = \frac{4}{3} \times \left\{ \begin{array}{ll}
 - \ln z + \frac{1}{3}~ ,\quad \quad \quad \quad z \rightarrow 0~, \\
~\left( \ln z  + const \right) \frac{1}{z} ~,\quad z \rightarrow \infty~.
\end{array} \right.  \end{equation}
Finally, calculating with the found function the integral in Eq.(\ref{F}) we
derive the e-line radiative correction (\ref{delta})

\begin{equation}\label{final}
\delta_{\mathrm e-line}^{\mathrm PNC} = - 1.97 \, \alpha^2 Z~.
\end{equation}
For heavy atoms the found correction is negative and
large.  This qualitative result agrees with Ref.\cite{kf_se_02}.  Moreover,
numerical results for the Cs atom are also in good agreement.
Eq.(\ref{final}) predicts that the e-line correction is $-0.6
\%$, which is close to  $-0.7(2) \% $ found in \cite{kf_se_02}.

Eq.(\ref{final}) shows that the e-line correction is large due to a large
coefficient $\sim 2.0$ in its right-hand side. {\em Naively} one could
expect this coefficient to be smaller, of the order of $ \sim 1/\pi$. It is
interesting that similar ``numerical enhancement'' happens for the e-line
radiative correction for the energy shift that is due to the finite nuclear
size (FNS).  Analytically this correction was examined by Pachucki
\cite{pachucki_93} and Eides and Grotch \cite{eides_97}.  In our notation
their result can be written as

\begin{equation}\label{FNS}
\delta_{\mathrm e-line}^{\mathrm FNS} =
{\cal C}^{\mathrm FNS} \alpha^2 Z = 2.978 \,\alpha^2 Z~,
\end{equation}
where ${\cal C}^{\mathrm FNS} = (3/2) \,1.985=2.978 $, the coefficient
1.985 is taken from Eq.(9) of \cite{eides_97}.  Eq.(\ref{FNS}) shows, that,
indeed, the e-line corrections to the FNS energy shift are governed by the
large coefficient $\sim 3.0$, similar to Eq.(\ref{final}) for the corrections
to the PNC amplitude. Fig.2 examines this similarity in more detail. It
shows data available for relative e-line corrections for the two problems
mentioned above, namely for the PNC amplitude and FNS energy shift. The
linear in $Z$ approximations (\ref{final}),(\ref{FNS}) 
(that are valid for sufficiently small values of $Z$)
are compared in
this figure with results of numerical calculations available for large
$Z$.  The e-line FNS corrections were calculated in
Refs.\cite{johnson_soff_85,blundell_FNS_92,cheng_93,lindgren_93}.
Specifically, the results for FNS shown in Fig.2 were
extracted in \cite{kf_se_02} from \cite{cheng_93} (using also
Ref.\cite{cheng_02} for $Z=55$), see details in
\cite{kf_se_02}. The data shown for the PNC amplitude are taken from the same
paper \cite{kf_se_02} that explains that the correction to the PNC
amplitude is equal to the average of the corrections to the
FNS energy shifts  for
$S_{1/2}$ and $P_{1/2}$ levels. 
The latter ones (i.e. the corrections to the FNS energy shifts of
$P_{1/2}$ levels) were also calculated in
Refs.\cite{johnson_soff_85,blundell_FNS_92,cheng_93,lindgren_93}.  The
results for the PNC amplitude shown in Fig.2 are based on data of
\cite{cheng_93} (and \cite{cheng_02}), 
see \cite{kf_se_02} for details.  We observe very close
quantitative similarity between corrections to PNC and FNS. 
In both cases the linear approximations (\ref{final}) and
(\ref{FNS}) predict large negative
corrections, which agrees with results based on numerical 
calculations for heavy atoms. 
Numerical validity of the linear approximations seem to be 
limited by the region below  $Z = 55$, 
for higher $Z$ they underestimate the effect.  
The Cs atom lies on the border, where results of
small-$Z$ and large-$Z$ approaches agree reasonably well.

Numerical data used in Ref.\cite{kf_se_02}
incorporates an error that increases 
for smaller values of $Z$, 
see Tables III and IV of \cite{cheng_93}. 
In order to reduce an impact of this error
we can combine Eq.(\ref{final}) of the present paper
with results of \cite{kf_se_02}.
Let us approximate the nonlinear terms omitted in Eq.(\ref{final})
by a simplest quadratic term and choose a corresponding coefficient 
to reproduce the results 
based on numerical data for very large $Z$, $Z \sim 90$, 
where the numerical errors are small. 
This gives us  the following 
interpolating formula for the corrections 
to the PNC amplitude

\begin{equation}\label{interp}
\delta_{\mathrm e-line,~int}^{\mathrm PNC} = - 1.97 \, \alpha^2 Z
(1+1.55 \,\alpha Z)~.
\end{equation}
Fig.2 shows that {\em any} reasonable
interpolation between data available for
large-Z and small-Z regions would produce a similar pattern.
We conclude therefore that Eq.(\ref{interp}) should give 
the most reliable numerical data. For the Cs atom we find from it 
the correction $-0.9(2) \% $, where we adopt the error
$0.2 \% $ of \cite{kf_se_02} (compare this result with
$-0.6 \%$ of Eq.(\ref{final}) 
and $ -0.7(2) \% $ of Ref.\cite{kf_se_02} mentioned above). 

The standard model value for the nuclear weak charge for Cs
\cite{groom_00} is

\begin{equation}\label{QW}
Q_W(^{133}{\mathrm Cs}) = \,-73.09 \,\pm\,(0.03)~.
\end{equation}
Ref. \cite{dzuba_02} refined previous calculations of Ref. \cite{dzuba_89}
extracting from the experimental PNC amplitude of Ref. \cite{wood_97}
the weak charge

\begin{equation}\label{72.18}
Q_W(^{133}{\mathrm Cs}) = \,-72.18\pm(0.29)_{\mathrm expt}\pm(0.36)_{\mathrm
theor}~,
\end{equation}
with the theoretical error $0.5\%$. It is consistent with
$Q_W(^{133}{\mathrm Cs}) = \,-72.21 \,\pm\,(0.28)_{\mathrm expt}\,\pm\,
(0.34)_{\mathrm theor}$ that was adopted in \cite{johnson_01} by taking the
average of the results of Refs.\cite{dzuba_89,blundell_92,kozlov_01}, and
accepting the theoretical error $0.4\%$ of \cite{bennett_wieman_99}. The
weak charge in Eq.(\ref{72.18}) deviates from the standard model (\ref{QW})
by $2.0\sigma$.  Taking  from Eq.(\ref{interp}) the value  $-0.9(2) \%$
for the e-line radiative correction, we derive from Eq.(\ref{72.18})

\begin{equation}\label{72.83}
Q_W(^{133}{\mathrm Cs}) =  \,-72.83 \pm(0.29)_{\mathrm
expt}\pm(0.39)_{\mathrm theor}~, \end{equation} which brings the
experimental results of \cite{wood_97} within the limits of the standard
model (\ref{QW}). (Note that even the smaller value for this correction
$-0.6 \% $ that follows from the linear approximation Eq.(\ref{final})
justifies the latter conclusion.)

The calculations presented show that the QED self-energy and vertex
radiative corrections bring the experimental results 
of Wood {\em et al} \cite{wood_97}
within the limits of the standard model.

This work was supported by the Australian Research Council. I wish to
thank Victor Flambaum for numerous discussions, and K.T.Cheng 
for Ref.\cite{cheng_02}.

\appendix

\section{}

The calculations performed use the following algebraic relations

\begin{eqnarray}\label{iden}
&& \langle \gamma_5 \rangle = 0~, \quad
\langle \gamma_\mu \gamma_5 \rangle = -a_\mu~, \quad
\langle \gamma_\mu \gamma_\nu \gamma_5 \rangle =
-(p_\mu a_\nu-p_\nu a_\mu) ~,
\\ \nonumber
&& \langle \gamma_\mu \gamma_\nu \gamma_\lambda \gamma_5 \rangle =
-(a_\mu g_{\nu\lambda} - a_\nu g_{\mu \lambda} + a_\lambda g_{\mu\nu} )
-i \epsilon_{\mu\nu\lambda\rho}a^\rho~,
\end{eqnarray}
where the $\langle X \rangle \equiv {\mathrm Tr}(X\wp)$,
and conventional integrals

\begin{eqnarray}\label{integrals}
\int \frac{d^4q}{(q^2+2qs-M^2)^n} &=&
\frac{(-1)^{n}\,i \pi^2 c_n}{(M^2 + s^2)^{n-2} }~,
\quad n>2~,
\\ \nonumber
\int \frac{q^\mu \,\,d^4q}{(q^2+2qs-M^2)^n} &=&
\frac{(-1)^{n-1} \,i \pi^2 c_n\, s^\mu~}{ (M^2 +s^2)^{n-2} },
\quad n>2~,
\\ \nonumber
\int \frac{q^\mu q^\nu \,\,d^4q}{(q^2+2qs-M^2)^3}
&=& \frac{-i\pi^2}{2}\,\left[ \, \frac{s^\mu s^\nu}{M^2 +s^2} -
\frac{1}{2}\,g^{\mu\nu}\left( \ln
\frac{\Lambda^2}{M^2+s^2}-\frac{3}{2}\right) \, \right ],
\\ \nonumber
\int \frac{q^\mu q^\nu q^\lambda \,\,d^4q}{(q^2+2qs-M^2)^n} &=&
\frac{(-1)^{n} \,i \pi^2 c_n}{(M^2+s^2)^{n-3}}\,\left[ -
\frac{s^\mu s^\nu s^\lambda}{M^2 + s^2} +
\frac{ g^{\mu\nu} s^\lambda + g^{\mu\lambda} s^\nu +
g^{\nu\lambda} s^\mu} {2(n-3)}\right], ~ n>3,
\end{eqnarray}
where $c_n= 1/[(n-1)(n-2)]$, $M$ is a scalar, $s$ is a
four-vector, $qs \equiv (qs)$, and $\Lambda$ is the ultraviolet cut-off.

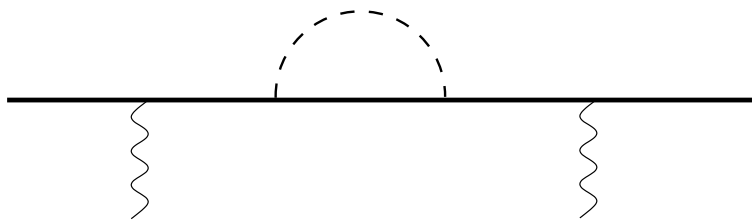
\begin{figure}
\caption{\label{one} The QED self-energy and vertex corrections, called
e-line corrections in the text, to the PNC matrix element. For each
diagram one of the wavy legs shows the Coulomb interaction with the
nucleus, another one the weak PNC interaction with the nucleus. Each
diagram represents all possible Feynman diagramms with the given
topological structure.} \end{figure}

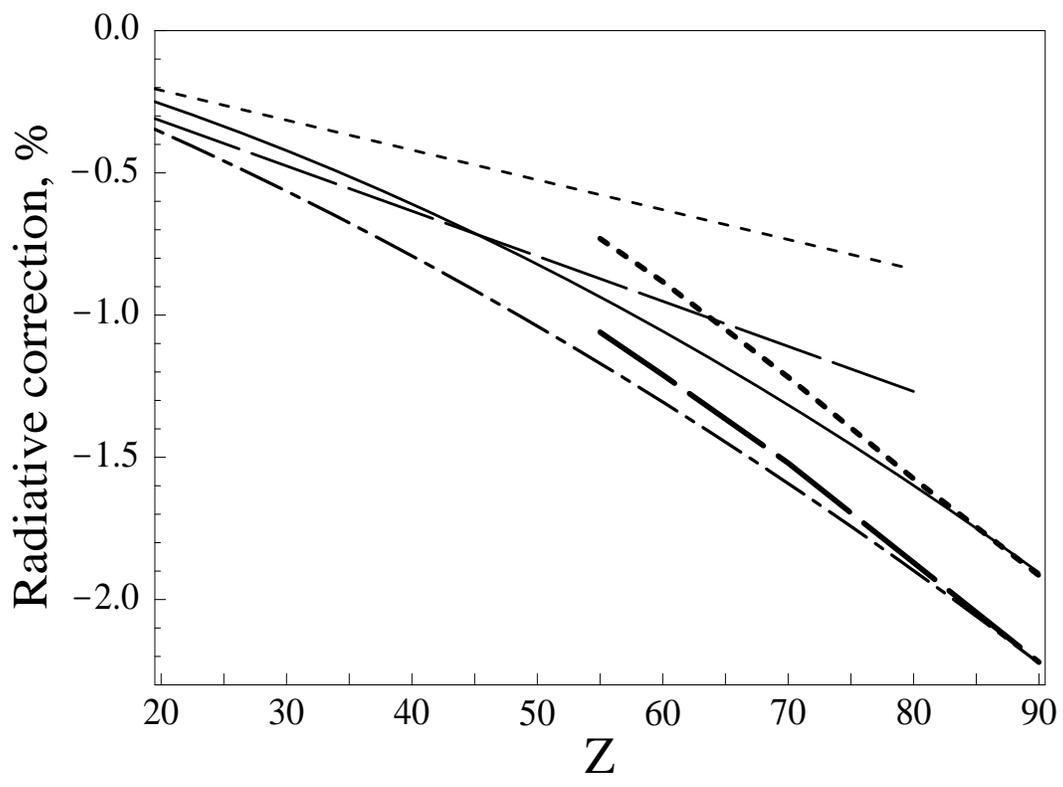
\begin{figure}
\caption{\label{two}
The thin dotted line, thick dotted line, and solid line:
e-line corrections to the PNC amplitude that follow from
Eq.(\ref{final}), Ref.[18], and Eq.(\ref{interp}).
The thin dashed line and thick dashed line:
the e-line corrections to the FNS energy shifts
that follow from Eq.(\ref{FNS}) (derived from [25]),
and Ref.[18]  (based on calculations of [22]), the dashed-dotted line:
the interpolation  
$\delta_{\mathrm e-line,~int}^{\mathrm FNS} = - 2.978 \, \alpha^2 Z
(1+0.85 \,\alpha Z)$ between the two latter lines.}
\end{figure}
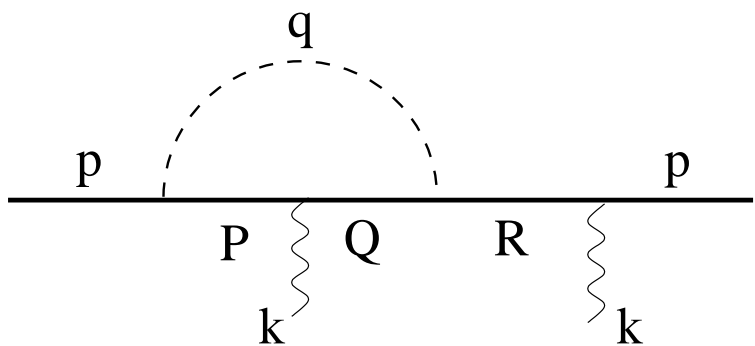
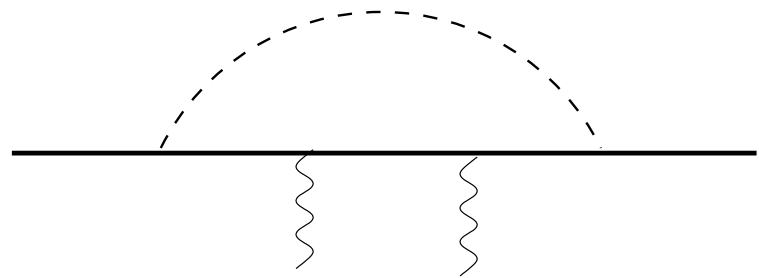
\end{document}